\documentclass[runningheads]{llncs}
\usepackage[T1]{fontenc}
\usepackage{graphicx}

\newcommand{\lo}{\mbox{LO}}
\newcommand{\lokit}{\mbox{LoKit}}
\newcommand{\myrule}[3]{
   {\noindent \small (#1) {\em #2} \nopagebreak \newline #3 \medskip} }
\newcommand{\becomes}   {\newline $\rightarrow$ \ }

\newcommand{\lpar}      {\newline @ \ }
\newcommand{\with}      {\newline \& \ }
\newcommand{\terminate} {\#t}

\newcommand{\bcast}     {{\bf \^{}}}

\newcommand{\tuple}[1]{{\em #1\/}}

\newcommand{\msgrequest}             {{\bf msg\_request}}
\newcommand{\msgquery}               {{\bf msg\_query}}
\newcommand{\msgreply}               {{\bf msg\_reply}}
\newcommand{\timeout}                {{\bf timeout}}
\newcommand{\blocked}                {{\bf blocked}}
\newcommand{\blockedotherserver}     {{\bf blocked\_other\_server}}
\newcommand{\catchfollowupreplies}   {{\bf catch\_follow-up}}
\newcommand{\serverS}                {{\bf server}(\tuple{S})}
\newcommand{\clientC}                {{\bf client}(\tuple{C})}
\newcommand{\client}                 {{\bf client}}
\newcommand{\timer}                  {{\bf timer}}

\newcommand{\replynolast}            {{\bf replyno\_last}}
\newcommand{\Id}              {\tuple{Id}}
\newcommand{\T}               {\tuple{T}}
\newcommand{\C}               {\tuple{C}}
\newcommand{\varS}            {\tuple{S}}
\newcommand{\A}               {\tuple{A}}
\newcommand{\Aeins}           {\tuple{A$_1$}}
\newcommand{\Azwei}           {\tuple{A$_2$}}
\newcommand{\B}               {\tuple{B}}
\newcommand{\Beins}           {\tuple{B$_1$}}
\newcommand{\Bzwei}           {\tuple{B$_2$}}
\newcommand{\Am}              {\tuple{Am}}
\newcommand{\trigger}         {\tuple{some\_trigger}}

\newcommand{\result}          {\tuple{result}}
\newcommand{\serverid}        {\tuple{server\_Id}}
\newcommand{\firstserverid}   {\tuple{first\_server\_Id}}
\newcommand{\otherserverid}   {\tuple{other\_server\_Id}}
\newcommand{\replyno}         {\tuple{replyno}}
\newcommand{\replynosucc}     {\tuple{replyno} + 1}
\newcommand{\rpcid}           {\tuple{RPC\_Id}}
\newcommand{\queryid}         {\tuple{query\_Id}}
\newcommand{\paramsN}         {\tuple{param$_0$},$\ldots$,\tuple{param$_N$}}
\newcommand{\somepolicy}          {{\sc application-dependent policy for timeout}}
\newcommand{\wait}                {{\sc wait}}
\newcommand{\produce}             {{\sc produce}}
\newcommand{\consume}             {{\sc consume}}
\newcommand{\getuniqueRPCid}      {{\sc get\_unique}(\rpcid)}
\newcommand{\getuniqueQueryid}    {{\sc get\_unique}(\queryid)}
\newcommand{\lookup}              {{\sc lookup}}
\newcommand{\deposit}             {{\sc deposit}}
\newcommand{\withdraw}            {{\sc withdraw}}
\newcommand{\transfer}            {{\sc transfer}}
\newcommand{\create}              {{\sc create}}
\newcommand{\delete}              {{\sc delete}}
\newcommand{\resume}              {{\sc resume}}
\newcommand{\suspend}             {{\sc suspend}}

\begin{document}

\title{\lokit\ (revisited): A Toolkit for Building Distributed Collaborative Applications}
\titlerunning{\lokit\ (revisited)}

\author{Uwe M. Borghoff}
\authorrunning{Uwe M. Borghoff}

\institute{Institute for Software Technology \\ 
University of the Bundeswehr Munich,  Neubiberg, Germany \\
\email{uwe.borghoff@unibw.de}}

\maketitle

\begin{abstract}
\lokit\ is a toolkit based on the coordination language \lo.
It allows to build distributed collaborative applications by providing a set of generic tools.

This paper briefly introduces the concept of the toolkit,
presents a subset of the \lokit\ tools, and finally demonstrates its
power by discussing a sample application built with the toolkit.

\keywords{
\lo\ \and fault-tolerant communication \and coordination \and toolkit.}

\end{abstract}

\section{Introduction}
This paper\footnote{This paper is a reprint of an unpublished report on the occasion of the (fictitious) 30th anniversary of the \textsc{Xerox Research Centre Europe}, now \textsc{Naver Labs}, Grenoble, France (https://europe.naverlabs.com/). \lokit\ was the basis for the development of the Constraint-Based Knowledge Brokers \cite{pasco/AndreoliBP94,jsc/AndreoliBP96,jucs/AndreoliBPS95}, a new type of search engine architecture at the time, which was able to search heterogeneous and multilingual databases in the same way \cite{paam/BorghoffPKNS96,jucs/BorghoffS96,riao/ChidlovskiiBC97} and establish cross-relationships between them via so-called Signed Feature Constraints \cite{pact2/AndreoliBP97,scp/BorghoffPAF98,ercimdl/ChidlovskiiB98}. This basic technology was marketed by \textsc{Xerox} from around 1999 under the name \textsc{Ask Once} and acquired by \textsc{Documentum} in 2004.} presents the basic design decisions and first
results for a  toolkit residing on top of the
coordination language \lo. 

\subsubsection*{\em The underlying language.}
\lo\ is a rule-based coordination language designed for managing
multiple software agents in open systems.  \lo\ can be applied for
implementing systems whose behavior is generated by the interaction of
autonomous, heterogeneous components.
In general, with \lo\ you can flexibly
customize and efficiently execute complex multi-agent coordination
schemata.
It is important to understand in which sense \lo\ 
-- and therefore the tools provided -- 
is a {\em coordination\/}
language, versus a programming language in the traditional sense.
This essentially means that \lo\ is just concerned about the coordination
and communication of events among agents, while traditional
``procedural'' activities (data-structure manipulation, arithmetics etc.)
are delegated to the programs embedded in the agents themselves.
Nowadays a language like \lo\ is, therefore, called {\em middleware}.

The basic features of the language are the @-construct for tight
coordination, the \&-construct for loose coordination, and the
\^{}-construct for broadcast communication.  The main uses of these
constructs are as follows.
Firstly, to inform the agents about actions being performed within
the distributed environment (monitoring, etc.),
secondly, to coordinate the actions performed by a {\em single\/} agent and, 
finally, to support actions involving {\em multiple\/} agents,
such as allocating/replicating resources among different nodes in the
network \cite{arcs/Borghoff90,icdt/Borghoff90}, 
or archiving/retrieving geographically dispersed information.

The @-construct for tight coordination is called {\em par\/}
because it assumes that all resources on the left-hand side
of a rule are consumed in parallel under
the all-or-nothing principle and that all resources on the right-hand side
are produced.
The \&-construct for loose coordination is called {\em with\/} because
it assumes that all resources at the left hand side of a rule
are consumed in parallel under the all-or-nothing principle and that all
resources on the right hand side of a rule separated by '\&'
are produced as replicas that act in parallel.
The \^{}-construct for broadcast communication is realized over
any transport layer that provides communication between --
if necessary -- heterogeneous sites.

It is beyond of scope of this paper to explain the full functionality and
semantics of the \lo-language;
for more details see \cite{Andreoli1992,Andreoli12,Andreoli4,Andreoli8}.

\subsubsection*{\em The toolkit.}
The aim of most CSCW-toolkits for building distributed applications
is to reduce development effort, to enable rapid prototyping, and
to increase the product quality of collaborative applications, e.g.,
GroupKit \cite{Roseman1992} for real-time conferencing,
Strudel \cite{Shepherd1990} for email-based workflow management,
or DistEdit \cite{Knister1990} for distributed editing.

The aim of \lokit\ is to provide coordination between
applications rather than providing applications themselves.
The design of \lokit\ is based on the following two main toolsets.
Firstly, the basic coordination toolset is of 
interest where coordination is achieved both between
multiple applications on a single host and
between multiple applications that run on different hosts.
The operating systems used at the different hosts (multi-vendor PCs and 
workstations) may range from PC-based Windows to NT, UNIXes or OS/2. All
communication is done within a client-server model, i.e., there are
clients (typically PCs) requesting tasks that involve coordination 
and servers that provide the tasks in question. 

Secondly, the higher level toolset is an important issue. \lokit\ provides
tools for system support such as different styles of communication,
failure handling, and transactions.
As far as the different communication styles are concerned,
the tools provide modules for RPC-like communication, 
one-way message-passing, both blocking (like CSP \cite{Hoare1985}) and 
non-blocking (like Actor-systems \cite{Agha1986}). Furthermore, it handles
queries with multiple replies and finally, fully-symmetric communication.
In this way, instead of providing simple transport-level
messaging (like send-message and receive-message primitives),
the toolkit provides high-level mechanisms that are more
appropriate for CSCW-applications.
The full set of higher level tools is generic 
and provides an easy way to implement different applications.

\subsubsection*{\em Organization of the paper.}
Sect.~\ref{primitives} starts with a discussion of two styles of
communication the toolkit provides, namely RPC-style and query-style.
The discussion includes a description of the generic rules
for both non-replicated and replicated servers, where the
number of replicas may a priori be unknown (e.g., during a WAIS lookup).
Moreover, rule sets are introduced that can handle timeouts within the
declarative \lo-environment of the toolkit.
In Sect.~\ref{banking} an example for remote banking is briefly discussed
that serves as a real-life application to demonstrate both
the appropriateness of the model and the usability of \lo\ and the toolkit
itself. 
Finally, Sect.~\ref{future} addresses topics for further research.

\section{\lokit-primitives for communication} \label{primitives}
\begin{figure*}[htb]
\includegraphics[width=\textwidth]{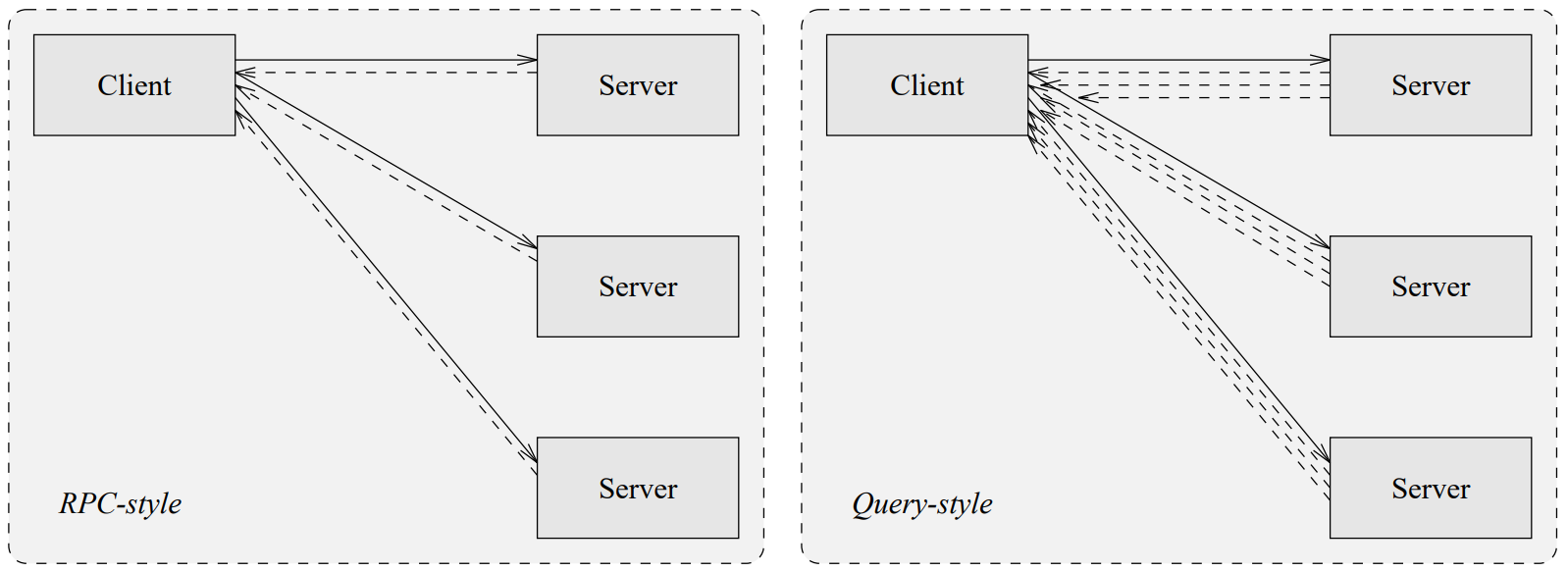}
\caption{Two example styles of communication.}
\label{commstyles}
\end{figure*}

Most of the applications within the area of interest need some form of
communication among distributed processes.
We provide amongst others, the basic mechanisms for RPCs and queries.
Fig.~\ref{commstyles} illustrates RPC-style and query-style forms
of communication with a single client and replicated servers.
Multiple replicas of the servers are installed to implement a
reliable service (see \cite{Wood1993} for a replicated RPC library
for the Amoeba distributed operating system).

For each of the these communication styles we 
discuss a possible \lo-realisation in terms of generic rules and
coordination features provided by the language. 
We motivate the decisions and give some remarks for further
extensions such as availability policy.

Within the description of the generic rules sometimes there is a need
for system-calls, system-dependent calculations, or application-dependent 
executions. {\em Resources\/} of this kind and real triggers
are printed in small capital letters, e.g.,
\getuniqueRPCid\ or \lookup.
Keywords of the generic rules are printed in bold face (e.g., \client), 
parameters and variables are printed in italic (e.g., \C).

\subsection{RPC-style connection}

An RPC-style connection is a synchronous communication, where
a client and a server together fulfill a certain task.\footnote{We are well-aware
of the existence of deferred synchronous communication
and asynchronous RPCs where placeholders for the return information
(e.g., the {\em futures\/} in {\em Cronus\/} \cite{Dean87,Gurw86}
or Liskov's and Shrira's {\em promises\/} \cite{Liskov1988})
are used and where the client may proceed until the return values
are really needed. Ananda {\em et~al.} give a comprehensive survey 
of asynchronous RPCs in \cite{Ananda1992}.
}
The communication is initiated by the client. It is a single outbound
flow (typically a request) with a single response/re\-ply (see a single
client-server-relation in Fig.~\ref{commstyles} -- RPC-style).

In the following, we show the \lo-level coordination part of the 
RPC communication, both
with and without replication of servers.

\subsubsection*{\em RPC without replication of servers.}
Under the assumption that there is only one provider (i.e., a single
server) for a particular service, the rules for connecting
a client with a server are quite simple. In fact, the client
simply sends out a request to all possible locations where the server
could reside. It then waits for the reply to arrive as a message
from a server.
To make things simple,
for the time being, we do not discuss time, i.e., here we simply
assume that the reply to the request will eventually arrive.
Later, of course, the client will
use some timeout mechanism to cover the case where
a reply is never received.

\myrule{R1}{client: initiate RPC-request}{
\clientC 
\lpar \getuniqueRPCid
\lpar \trigger
\lpar \bcast \msgrequest(\rpcid,\paramsN)
\becomes \client(\C,\blocked,\rpcid).
}

\myrule{R2}{server: produce result and reply to RPC-request}{
\serverS 
\lpar \msgrequest(\rpcid,\paramsN)
\lpar \produce(\rpcid,\paramsN,\result)
\lpar \bcast \msgreply(\serverS,\rpcid,\result)
\becomes 
\serverS.
}

\myrule{R3}{client: consume RPC-reply}{
\client(\C,\blocked,\rpcid)
\lpar \msgreply(\serverid,\rpcid,\result)
\lpar \consume(\serverid,\rpcid,\result)
\becomes 
\clientC.
}

(R1) The client initiates the RPC-style connection when the trigger is
consumable, e.g., a trigger-event from the user interface or
a particular trigger message from outside.
The client sends the request to the possible locations of the server that
should handle the request. 
The RPC-request itself has an unique identifier (\rpcid) in order to
connect the request-reply activity, i.e., it serves for the
communication binding.
The unique identifier is created through the system-level
command \getuniqueRPCid.
Furthermore, the request carries $N$+1 parameters to fulfill the request
at the server side.
In a RPC-style connection, parameter \tuple{param$_0$} is typically the name
of the procedure to be invoked.

(R2) The corresponding server is triggered by the
request message.
It carries out the request by producing a result that is sent
back in a reply message
to the initiator of the request.
The initiator of the request is already
in blocked mode (\client(\C,\blocked,\rpcid)). In the current
implementation the production of the result is done in 
a low-level Prolog implementation that depends on the
task to fulfill.

(R3) The blocked client that is triggered 
by the reply message with the corresponding
\rpcid\ consumes the result offered by the server \serverid.

\subsubsection*{\em RPC with replication of servers.}
Under the assumption that there are possibly more providers (i.e., multiple
identical servers)
for a particular service, the rules for connecting
a client with these servers become more complicated. The client
sends out a request to all possible locations where the servers
could reside. It then waits for the replies to arrive as different messages
from these servers.

In many applications, the number of servers are a priori unknown.
After starting, for instance a world-wide WAIS search, the number of servers
contacted and willing to reply cannot be predicted.
Since it is not realistic and sometimes impossible
to wait for the replies of all servers, 
we provide different
instances of the client w.r.t.\ answers received.
First of all there is the ``main'' client that waits for the first
reply to arrive. As soon as the first reply has been consumed,
this client may proceed (as in the implementation shown below).

In parallel, this client creates another instance of itself, the
so-called {\em catch-follow-up\/} client. This catch-follow-up client, and all further
catch-follow-up clients that are created in the same way, will process the
other replies from the ``slower'' servers.

\myrule{Rr3}{client: consume RPC-reply and spawn catch-follow-up client}{
\client(\C,\blocked,\rpcid)
\lpar \msgreply(\serverid,\rpcid,\result)
\lpar \consume(\serverid,\rpcid,\result)
\becomes 
\clientC\ 
\with
\client(\C,\catchfollowupreplies,\rpcid).
}

\myrule{Rr4}{catch-follow-up client: consume other RPC-reply and spawn new catch-follow-up client}{
\client(\C,\catchfollowupreplies,\rpcid)
\lpar \msgreply(\serverid,\rpcid,\result)
\lpar \consume(\serverid,\rpcid,\result)
\becomes 
\client(\C,\catchfollowupreplies,\rpcid).
}

(Rr1) As in the case with no replication of servers, 
the client initiates the RPC-style connection when the trigger is
consumable and blocks.
The client broadcasts the request to all possible servers that
are able to handle the request. 
Note, however, that all of them get a copy of the request.

(Rr2) For each of the servers, as before in the non-replicated case.

(Rr3) The blocked client that is triggered 
by the reply message with the corresponding
\rpcid\ consumes the {first} result offered by one of the servers.
Furthermore, another client process is spawned that consumes (e.g.,
by deleting them) the {follow-up server-replies}
of the \rpcid-request that is already handled. 

(Rr4) For every single RPC there exists 
such a client process that consumes
the follow-up
server-replies of the \rpcid-request that is already handled. 
We have not provided a termination policy for these processes yet.
In fact, we do not know how many replies will be received 
for a single request
because we do not want to count active servers (this is part of
a possible availability policy). 
Furthermore, we do not know
how long at takes until the last reply is received.
Again at this level, we do not want to install a time-out
scheme. This should be part of the availability policy as well.
During the lifetime of our system
a lot of processes will be created (as a matter of fact, resources) 
of the form
\client(\C,\catchfollowupreplies,\rpcid).
To get rid of them, a garbage collector in every pool can be
defined that deletes such ``processes'' that are waiting for
RPC-replies of obsolete requests. 

Another, more elegant, possibility is implemented using
our timeout policy. Here the catch-follow-up clients
are terminated (\lo-notation: \terminate)
without any intervention from administrators
simply using a side-effect. 
More about this in Sect.~\ref{timer}.

\subsection{Query-style connection}

A query-style connection is a synchronous connection
between a client and some server.
The client sends out a single outbound flow of data (the query-request)
and receives multiple replies in a chain (see a single
client-server-relation in Fig.~\ref{commstyles} -- query-style).

\subsubsection*{\em Query without replication of servers.}
Assuming that there is only one server
for a particular service, the client
simply sends out a request to all possible locations where the server
could reside. It then waits for the reply to arrive as 
{\em multiple\/} messages from this server.

\myrule{Q1}{client: initiate query-request}{
\clientC 
\lpar \getuniqueQueryid
\lpar \trigger
\lpar \bcast \msgquery(\queryid,\paramsN)
\becomes \client(\C,\blocked(1),\queryid).
}

\myrule{Q2}{server: produce \replyno -th result and reply to query-request}{
{\hfill -- set of rules, not shown -- \hfill}
}

\myrule{Q3}{client: consume \replyno -th reply}{
\client(\C,\blocked(\replyno),\queryid)
\lpar \msgreply(\serverid,\replyno,\queryid,\result)
\lpar \consume(\serverid,\queryid,\result)
\becomes 
\client(\C,\blocked(\replynosucc),\queryid).
}

\myrule{Q4}{client: consume \replyno -th and last reply}{
\client(\C,\blocked(\replyno),\queryid)
\lpar \msgreply(\serverid,\replynolast,\replyno,\queryid,\result)
\lpar \consume(\serverid,\queryid,\result)
\becomes 
\clientC.
}

(Q1) The client initiates the query-style connection when the trigger is
consumable and blocks for the first reply to be received.
As in RPC-style connection there is a strict communication binding.
The query itself has an unique identifier (\queryid) to
connect the different request-reply activities to one task.
The unique identifier is created by the system-level
command \getuniqueQueryid.
Again, the request carries $N$+1 parameters to fulfill the request
at the server side where
parameter \tuple{param$_0$} is the name
of the query procedure to be invoked.

(Q2) Analogously to the server site rule in the RPC-cases,
a corresponding server is triggered by the query message.
It carries out the request by producing multiple results that are sent
back in a several reply messages to the initiator of the query.
Since \lo\ is an asynchronous coordination language, we cannot
guarantee that all replies of a server are received in order
(even when the network protocol used provides this feature).
Every reply message, therefore, is tagged with an unique \replyno\ ranging
from 1 to $n$, e.g.,
\msgreply(\serverS,1,\queryid,\result)
for the first reply message.
The last message is additionally tagged with \replynolast, i.e.,
the last message out of $n$ messages that are sent back to the client
has the following form:
\msgreply(\serverS,\replynolast,$n$,\queryid,\result).

(Q3) The blocked client that is triggered 
by the \replyno -th reply message with the corresponding
\queryid\ consumes the \replyno -th result offered by the server.
Then the client blocks for the next reply to be received
by the server.

(Q4) If the last reply is received (indicated by \replynolast)
the client is unblocked and ready for further activities.

\subsubsection*{\em Query with replication of servers.}
The more interesting case is where servers are replicated. 
Here the client must identify which reply message belongs to which server
(still allowing for out-of-order receptions).
The client initiation and the replies of each server 
are as before in the case without server replication. 
However, the client site code for the consumption of the replies
has to be modified, as shown in the following.

\myrule{Qr3}{client: consume first reply of first server}{
\client(\C,\blocked(1),\queryid)
\lpar \msgreply(\firstserverid,1,\queryid,\result)
\lpar \consume(\firstserverid,\queryid,\result)
\becomes 
\client(\C,\blockedotherserver(1),\queryid)
\with \client(\C,\blocked(\firstserverid,2),\queryid).
}

\myrule{Qr4}{client: consume first and last reply of first server}{
\client(\C,\blocked(1),\queryid)
\lpar \msgreply(\firstserverid,\replynolast,1,\queryid,\result)
\lpar \consume(\firstserverid,\queryid,\result)
\becomes 
\clientC\ 
\with 
\client(\C,\blockedotherserver(1),\queryid).
}

\myrule{Qr5}{client: consume \replyno -th reply of first server}{
\client(\C,\blocked(\firstserverid,\replyno),\queryid)
\lpar \msgreply(\firstserverid,\replyno,\queryid,\result)
\lpar \consume(\firstserverid,\queryid,\result)
\becomes 
\client(\C,\blocked(\firstserverid,\replynosucc),\queryid).
}

\myrule{Qr6}{client: consume \replyno -th and last reply of first server}{
\client(\C,\blocked(\firstserverid,\replyno),\queryid)
\lpar \msgreply(\firstserverid,\replynolast,\replyno,\queryid,\result)
\lpar \consume(\firstserverid,\queryid,\result)
\becomes 
\clientC.
}

\myrule{Qr7}{client: consume first reply of another server}{
\client(\C,\blockedotherserver(1),\queryid)
\lpar \msgreply(\otherserverid,1,\queryid,\result)
\lpar \consume(\otherserverid,\queryid,\result)
\becomes 
\client(\C,\blockedotherserver(1),\queryid)
\with \client(\C,\blockedotherserver(\otherserverid,2),\queryid).
}

\myrule{Qr8}{client: consume first and last reply of another server}{
\client(\C,\blockedotherserver(1),\queryid)
\lpar \msgreply(\otherserverid,\replynolast,1,\queryid,\result)
\lpar \consume(\otherserverid,\queryid,\result)
\becomes 
\client(\C,\blockedotherserver(1),\queryid).
}

\myrule{Qr9}{client: consume \replyno -th reply of another server}{
\client(\C,\blockedotherserver(\otherserverid,\replyno),\queryid)
\lpar \msgreply(\otherserverid,\replyno,\queryid,\result)
\lpar \consume(\otherserverid,\queryid,\result)
\becomes 
\client(\C,\blockedotherserver(\otherserverid,\replynosucc),\queryid).
}

\myrule{Qr10}{client: consume \replyno -th and last reply of another server}{
\client(\C,\blockedotherserver(\otherserverid,\replyno),\queryid)
\lpar \msgreply(\otherserverid,\replynolast,\replyno,\queryid,\result)
\lpar \consume(\otherserverid,\queryid,\result)
\becomes 
\terminate.
}
 
The rules (Qr1) and (Qr2) are as before in the non-replicated case.

(Qr3) The blocked client that is triggered 
by the first reply message with the corresponding
\queryid\ consumes this first result offered by the first
server that answers.
Then the client blocks for the second reply to be received
from the same server.

(Qr4) If the first reply is also the 
last reply to be received (indicated through \replynolast)
the client is unblocked and ready for further activities.
In both cases, an additional client is created that is blocked to 
receive the replies of another server.

The rules (Qr5) and (Qr6) catch the follow-up replies of the first server.

The rules (Qr7) and (Qr8) catch the first reply of another server.
If the other server sends at once the last reply, there
is no need to unblock the client.
In both cases, simply 
an additional client is created that is blocked to 
receive the replies of any further servers.

The rules (Qr9) and (Qr10) catch the follow-up replies of another server.
As soon as the last reply arrives, this additionally created
client terminates.
Note that in all of these rules, that even when the last message 
\msgreply(\serverS,\replynolast,$n$,\queryid,\result)
arrives too early (i.e., the client has not yet received/consumed
all messages from $1\ldots n-1$) the client will not be unblocked.
The client is still in a state
\client(\C,\blocked(\firstserverid,$k$),\queryid) with $k \in \{1,\ldots,n-1\}$ 
and the last-reply handling rule cannot be satisfied yet.
The same holds of course for
\client(\C,\blockedotherserver(\otherserverid,\replyno),\queryid).
In other words, all replies are consumed in sequence.
It is worth mentioning that this scheme exploits the fact that unconsumed
messages are held by the runtime environment until they are explicitly consumed, i.e.,
until the message is a resource on the left-hand side of a rule which fires.

As in the RPC-case,
for every single query there exists a
client process that consumes
the follow-up
server-replies of the \queryid-request that is already handled. 
For the same reason as discussed before,
we have not shown a termination policy for these processes yet 
(the timer-controlled policy will manage this).
Again we do not know how many reply messages will be received 
for a single query and, more important, how many servers will
answer.

However, during the lifetime of our system
only one (additional) process per query will be created, in contrast 
to one per reply.
This is due to the fact that the last reply can be identified,
and the process can be terminated as shown in the rule (Qr10).

As far as problem-oriented programming is concerned, we have already
mentioned the need for replication.
The next step is to implement replication policies
within the toolkit in order to increase availability as well as
to realize performance gains.
In fact, a variety of distributed system approaches
use some sort of replication scheme. For a comprehensive
survey see \cite{Borghoff1992book}.

Amongst others, replication can be exploited in two 
different ways: Firstly, it can be used to
mask system failures such as node crashes, or secondly, it can be used to speed up
workflow throughput by sending out the same task to different
``agent''-replicas. As soon as the ``fastest agent'' has replied, the
next step in the workflow can be triggered.
All RPCs or queries are then sent to the pool of replicated agents that 
may fulfill
the required task in parallel and fully independently. 
Furthermore, the pool of replicated agents can be installed
on different hosts to increase availability.
If some of the agents have crashed or are 
not reachable over the network, the client can still receive
replies from one of the remaining available agents.

The degree of replication can be dynamically adjusted to special needs
or particular pattern of the network (current load, link failure ratio, etc.).
What is needed is only a simple decision making routine within the client.
We have shown how a client can react to replies from different 
servers to the same RPC-request or the same query.
Take for instance the rule (Qr4), i.e., the
first rule within a client to receive
the first (and last) reply of the ``fastest'' server of a query.
When, in this rule, the resource
\client(\C,\blockedotherserver(1),\queryid)
is deleted, a client is created 
that is only interested in the first server to answer,
no matter which server that is.
All redundant replies of other servers are then ignored, i.e.,
if we have a write-all-read-any scheme and
all but one server crash, the system can still read without
harm to the task to fulfill. 

\subsection{The notion of time} \label{timer}
Distributed applications need some notion of time.
Some applications may require the execution of a particular
function at a concrete timepoint 
(e.g., given by a resource {\em time\/}("12:02:34")), 
or within a given interval of time
(e.g., given by a resource {\em time\_interval\/}("12:00:00-12:05:00"))
or, most importantly, may react to the network delays or site crashes
through a timeout. Most policies within distributed communication
rely on timeout {signals},
e.g., given by a resource {\em timeout\/}(\T) where \T\ is a variable holding
the seconds to wait at least until the timeout should occur.
In this sense {\em timeout\/}(5) should imply a timeout in 5 seconds.
We say ``should imply a timeout in 5 seconds'' because of the following
\lo-characteristics:
since the underlying language \lo\ that is used to implement the toolkit
is asynchronous and does not provide real-time features, we must
handle timeouts in another way. For us, a timeout is simply an upper bound
that, when reached, will eventually trigger a timeout handling routine.

To be more precise, let's assume the following scenario:
a client sends out a RPC-request as given by the rules in the above
examples. Together with the request it sets a timeout value, i.e.,
a time interval that is the minimum period the client will wait
for responses. If the timeout occurs the client will 
evaluate the responses (or no responses at all) and take a decision
whether to proceed normally or to invoke some failure handling
routine.

In a replicated server environment, the client may use a policy that
guarantees the consistency of the server data through the so-called
{\em majority consensus}.\footnote{{\em Majority consensus\/} is just one of the
schemes to maintain consistency among replicated data.
Other schemes include {\em write-all-read-any},
{\em primary copy}, {\em quorum consensus}, or the flood of
{\em voting\/} approaches; see \cite{infospek/Borghoff91} for a survey.}
A request for update is then successful if and only if
the majority of servers have fulfilled the update.
As long as the majority of servers has not replied to the request, the client
must not commit the update. After the timeout, the client
checks whether or not the majority has been reached and takes a commit or abort
decision. Of course if all servers reply within the given amount of time
no timeout handling is done.

The timeout control, i.e., who is responsible for storing the set timeout
value \T\ and for signaling the timeout, can be handled within the toolkit in
two different ways.
Firstly, the client itself can store the timeout value \T\ and check (in a
busy-wait loop) whether or not the timeout condition is satisfied.
Secondly, the client can spawn a {\em timer agent\/} that is provided
with the timer value. The client sleeps (or accepts incoming messages)
as long as the timer
agent does not signal the timeout or, in our given example, all servers
have replied.

For the implementation both variants have more or less the same
behavior. Since \lo-agents never sleep (but, they always check the
resource space for possible satisfaction of a rule to fire) the signal
of the timer agent is a resource provided to the agent resource space
that may then trigger a timeout-handling-rule to fire.
This checking in the resource space for the timeout-resource
is as expensive as the busy-wait described in the first variant.

As a matter of fact, the provision of the timeout-resource is not
a signal as known e.g. from Unix. There, immediately when the runtime
priority allows it, the corresponding process is interrupted and in the
end set to run the timeout-routing.
In \lo\ the time for firing a rule cannot be predicted and depends
heavily on the number of active agents, number of rules, and other
system parameters that change dynamically.

Therefore, our timeout model is a 
simulation of the behavior of a ``normal'' timeout
in the sense that we wait at least $T+\Delta t$ sec., where
$\Delta t$ is positive and depending on the parameters just described.

Since, however, we do not intend to provide real-time features,
we believe that this timeout model satisfies the needs of
almost all network communication protocols that will be build using the toolkit.
Often it is even not necessary to have a real clock timer but an
incremental counter, as proposed by e.g. Lamport's algorithm \cite{Lamport1978}
or all voting algorithms we know of \cite{infospek/Borghoff91}.
The \lokit-solution concerning the notion of time and timeouts
is as follows. 

\subsubsection*{\em Timer rule.}

\myrule{T}{timer: main rule}{
\timer(\C,\Id,\T)
\lpar \wait(\T)
\lpar \bcast \timeout(\C,\Id)
\becomes 
\terminate.
}

(T) We install an agent, called {\em timer\/}, that plays the role of a clock.
The timer waits the given amount of time. Then it simply 
broadcast the timeout message with the given specification.
As usual, there is a binding between timer-invocation and timeout delivery.

\subsubsection*{\em Timer setting.}
To initiate
a timer-controlled communication, the first rule of our communication styles
is simply enhanced by the creation, i.e. the setting, of a timer.
The following rules {\em replace\/} the corresponding rules, e.g.,
(Qr1\_t) replaces rule (Qr1).

\myrule{R1\_t) and (Rr1\_t}{client: initiate RPC-request (timer-controlled)}{
\clientC 
\lpar \getuniqueRPCid
\lpar \trigger
\lpar \bcast \msgrequest(\rpcid,\paramsN)
\becomes \client(\C,\blocked,\rpcid)
\with \timer(\C,\rpcid,\T).
}

\myrule{Q1\_t) and (Qr1\_t}{client: initiate query-request (timer-controlled)}{
\clientC 
\lpar \getuniqueQueryid
\lpar \trigger
\lpar \bcast \msgquery(\queryid,\paramsN)
\becomes \client(\C,\blocked(1),\queryid)
\with \timer(\C,\queryid,\T).
}

(R1\_t) and (Q1\_t) The timer 
is ``created'' when a client sends out a request. Thereby, the
client specifies itself (\C), the request that should be timer-controlled
or better timeout-controlled (\Id), and, finally, the timeout value itself
(\T), i.e., the time the client is willing to wait for the reply before
it wants to be interrupted.
Consequently, all client instantiations, i.e., 
\begin{itemize}
\item
the clients waiting for the RPC-reply (with and without replication of servers),
\item
the clients waiting for the first reply of a query-request (with and without
replication), and
\item
the clients waiting for follow-up replies to a given query-requests
\end{itemize}
must have an additional rule that handles the timeout.
Depending on the kind of client instantiation, there are different reactions
to the timeout as shown in the following rules.

{\bf Remark:}
Since the timer is implemented on the client site, the rules on the
server sites (i.e., (R2), (Rr2), etc.) are not affected.

\subsubsection*{\em Timer-controlled RPC.}
To handle timeouts, the following rules must be {\em added\/} to the
corresponding rule sets, e.g., the rules (Rr3\_t) and (Rr3\_t) are added to the rule set
consisting of the rules (Rr1\_t),(Rr2),(Rr3),(Rr4).

\myrule{R3\_t) and (Rr3\_t}{client: consume RPC-reply (timeout)}{
\client(\C,\blocked,\rpcid)
\lpar \timeout(\C,\rpcid)
\lpar \somepolicy
\becomes 
\clientC.
}

\myrule{Rr4\_t}{catch-follow-up client: consume other RPC-reply (timeout)}{
\client(\C,\catchfollowupreplies,\rpcid)
\lpar \timeout(\C,\rpcid)
\lpar \somepolicy
\becomes 
\terminate.
}

In the non-replicated as well as
replicated case of an RPC, we re-install the client with the rules (R3\_t)
and (Rr3\_t), respectively.
In the replicated case we, furthermore,
terminate the catch-follow-up client in rule (Rr4\_t).
As a side-effect, we always terminate the last
catch-follow-up client that would otherwise run forever. 
This is a useful feature of our timeout policy. 

\subsubsection*{\em Timer-controlled query.}
The rules for timeout-handling within our query-style communication
are analogously formed as illustrated below.
They must also be {\em added\/} to the corresponding rule sets.

\myrule{Q3\_t) and (Qr3\_t}{client: consume first reply of first server (timeout)}{
\client(\C,\blocked(1),\queryid)
\lpar \timeout(\C,\queryid)
\lpar \somepolicy
\becomes 
\clientC.
}

\myrule{Q4\_t) and (Qr4\_t}{client: consume \replyno -th reply of first server (timeout)}{
\client(\C,\blocked(\firstserverid,\replyno),\queryid)
\lpar \timeout(\C,\queryid)
\lpar \somepolicy
\becomes 
\clientC.
}

\myrule{Qr5\_t}{client: consume first reply of another server (timeout)}{
\client(\C,\blockedotherserver(1),\queryid)
\lpar \timeout(\C,\queryid)
\lpar \somepolicy
\becomes 
\terminate.
}

\myrule{Qr6\_t}{client: consume \replyno -th reply of another server (timeout)}{
\client(\C,\blockedotherserver(\otherserverid,\replyno),\queryid)
\lpar \timeout(\C,\queryid)
\lpar \somepolicy
\becomes 
\terminate.
}

In the non-replicated as well as
in the replicated case of a query, we re-install the client with the rules 
(Q3\_t) and (Q4\_t) as well as (Qr3\_t) and (Qr4\_t), respectively.
Furthermore, in the replicated case we
terminate the unwanted catch-follow-up clients with the rules
(Qr5\_t) and (Qr6\_t).

\begin{figure*}[ht]
\includegraphics[width=\textwidth]{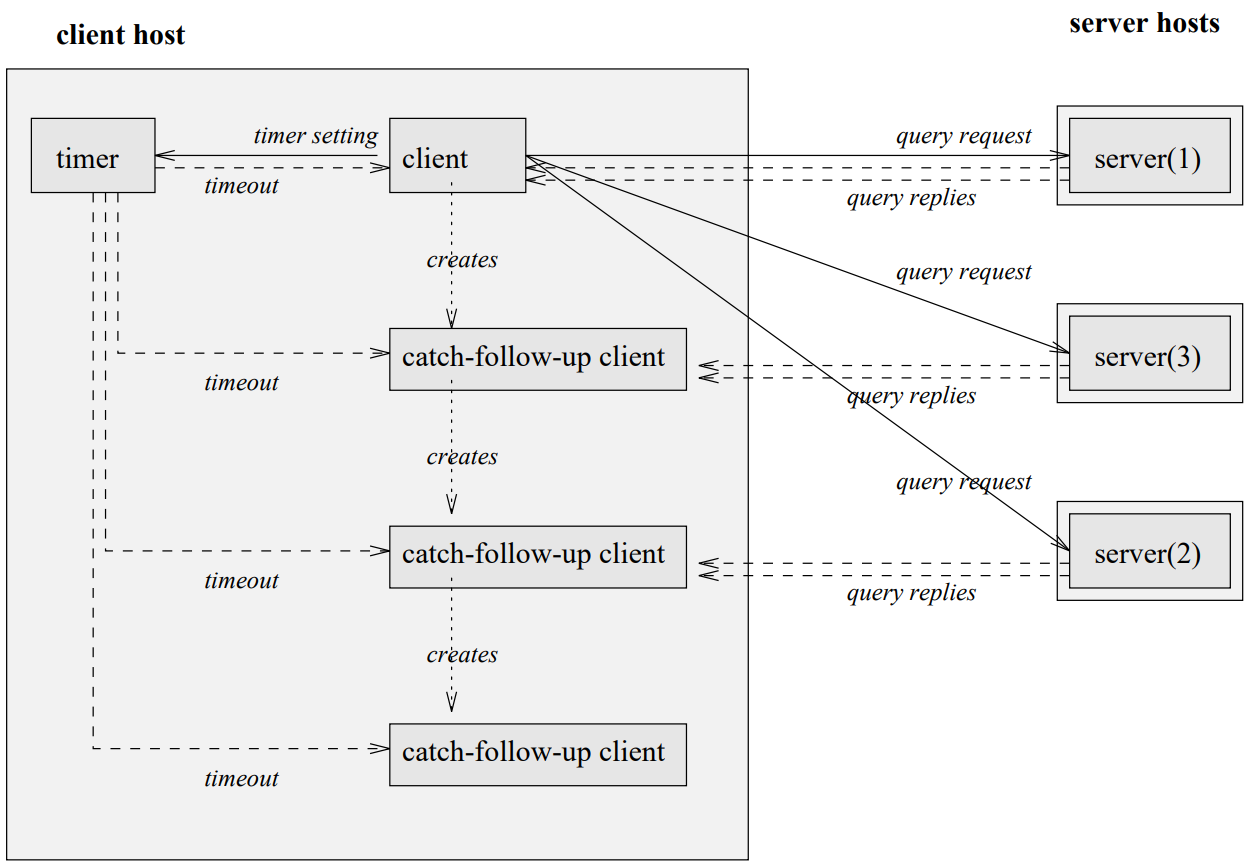}
  \caption{Timer-controlled query processing.}
  \label{query}
\end{figure*}

Fig.~\ref{query} illustrates the different roles during the timer-controlled
query processing.
In the given scenario three identical servers are installed. The query-request
has been sent to all of them. We assume that server(1) replies first, then server(3), and
finally, server(2). Every time the first reply message of a new server is received,
a catch-follow-up client is created. We assume, in addition, that every reply to the
query results in two separate reply messages that are sent back to the client host.
In the beginning, the client has set a timer. After the specified amount of time
the timer sends timeout messages to all client instantiations, i.e., to the
original client and all newly created catch-follow-up clients.

\section{Remote banking} \label{banking}
One of the project's application domains is in the area of remote banking.
Remote banking becomes an attractive domain and very effective
approach for fulfilling the potentialities that emerge from the following
observations: 
\begin{itemize}
\item electronic and geographically dispersed banking services
according to today's merchant requirements have the primary function
to provide easy access to, e.g., customer information accounts;
\item customers may process their requests from the customer sites
to the remote bank sites through public and corporate networks;
\item customers may have abstract interfaces for uniform access to the
basic banking services offered by different branches and for managing the
coordination-critical dialogue among such services;
\item corporate customers  may have direct access (e.g., for read-only
accesses or cash transfers within well-defined boundaries) 
to their current accounts without directly interacting with the
account owners, the banks.
\end{itemize}

Remote banking as our choice for a distributed application domain is 
furthermore motivated by the fact that it addresses a broad spectrum of 
users like end-users, subscribers of particular services,
basic service providers (BSP), and
value added service providers (VASP). It addresses, moreover,
a variety of state-of-the art technology like computing over wide area
networks. 
In its simpliest form, the VASP servers enable 
the creation of customized versions of remote banking services based on some criteria, e.g.,
input from customers and banks, available banking services such as deposits, withdrawals,
or money transfers,
corresponding requirements of the customers,
technology used by customers and banks such as operating system platforms and finally,
telecommunication idiosyncrasies.
Schael and Zeller describe in \cite{Schael1993} a banking workflow system
for particular financial services using a method based on a 
client-server model, or as they call it, client-supplier model.

The main focus lies on the design of a VASP which provides end-users and
subscribers of particular services with simplified access to the basic services offered
by the BSPs. Furthermore, the VASP should allow for composite services according to some
cost/revenue functions or some broker services.
These broker services are capable of dynamically selecting basic services w.r.t.\ 
user needs and user-given constraints.
Rapid customization should be achieved both on a subscriber basis (e.g.,
a department of a corporate customer) 
and on an end-user basis (e.g., employees in a department of a corporate customer).

To demonstrate the power of \lokit, we implemented a first prototype of
such a remote banking example with distributed replicated databases, i.e., 
each of a pre-defined number of servers is in charge of a replica of the database.
The consistency scheme
uses a {\em write-all-read-any\/} policy; replication transparency is provided. 
There may be multiple servers for each bank with separate processes running
on (perhaps) different hosts. Within our local area
network, the processes are connected through sockets.
Over a wide area network, we have already experimented with email as 
transport mechanism.
A client process that gets in contact with a bank 
(or in contact with more banks where appropriate) can trigger one 
of the tasks shown in Table~\ref{trigger}.

\begin{table*}[bt]
\small
\begin{center}
\begin{tabular}{| l | p{7.8cm} |}
\hline
\trigger & description of task \\
\hline \hline
\lookup &
shows all accounts of all banks.
\\
\lookup(\B) &
shows all accounts of bank \B.
\\
\lookup(\B,\A) &
shows account \A\ of bank \B.
\\ \hline
\deposit(\B,\A,\Am) &
adds amount \Am\ to account \A\ of bank \B.
\\
\withdraw(\B,\A,\Am) &
subtracts amount \Am\ from account \A\ of bank \B\
if and only if there is no overdraft at \A.
\\
\transfer(\Beins,\Aeins,\Am,\Bzwei,\Azwei) &
transfers amount \Am\ from account \Aeins\
of bank \Beins\ to account \Azwei\ of bank \Bzwei\
if and only if there is no overdraft at \Aeins.
\\ \hline
\create(\B,\A) &
creates account \A\ at bank \B\
if and only if this account does not exist already.
\\
\delete(\B,\A) &
deletes account \A\ at bank \B\
if and only if this account has balance 0.
\\
\hline
\end{tabular}
\end{center}

{\footnotesize
{\bf Remarks:} In the current implementation, \lookup\ starts a query whereas
all other triggers result in RPCs. 
In response to a \lookup, the corresponding
servers return the statement list of every account in a separate message.
A statement list includes the balance and the transaction history of the account
together with detailed information on attempts to overdraw the account.

\transfer(\Beins,\Aeins,\Am,\Bzwei,\Azwei) is part of a higher layer 
and is built on top of the two constructs
\deposit(\Bzwei,\Azwei,\Am) and
\withdraw(\Beins,\Aeins,\Am).
The transfer is {\em atomic\/} with nested transactional behavior \cite{Lynch1994}.

\ 
}
\caption{Possible tasks of a remote banking client.}
\label{trigger}
\end{table*}

During the development of the distributed coordination there is a need to test
the failure transparency of the proposed strategies, i.e., masking of
network and link failures. Therefore the toolkit provides means for
simulating these kind of failures.
As a first approach, network failures (especially partitionings) and
failstop site crashes \cite{Schl83} can be simulated using the additional triggers 
\suspend(\varS) or \resume(\varS) to suspend or
resume the execution of a server \varS, respectively.
The first results are impressive and show that \lokit\ could be used
for rapid prototyping and simulation of distributed applications.

The current implementation of the toolkit runs on Unix-workstations. Ports to other
platforms are planned. 

\section{Future work}  \label{future}
The following list summarizes the possible evolution of \lokit\ as
a basic toolkit for the use within distributed applications.
First of all, an interface to \lokit\ should be designed
that could be used for
rapid prototyping \lokit-based applications. Since the rules
provided by \lokit\ allow for safe and systematic programming,
this interface could play the role of a compiler that creates
\lokit\ rules that are then interpreted by the underlying
\lo-platform.
The interface could have the form of a template where all the 
application-specific parts of the rules 
(i.e., underlined resources in the rule set given before)
are typed in.
Together with explicitly formulated predicate-libraries
(for the time being, in Prolog), the template will be a sort of
graphical application programming interface.

\begin{figure*}[ht]
\includegraphics[width=\textwidth]{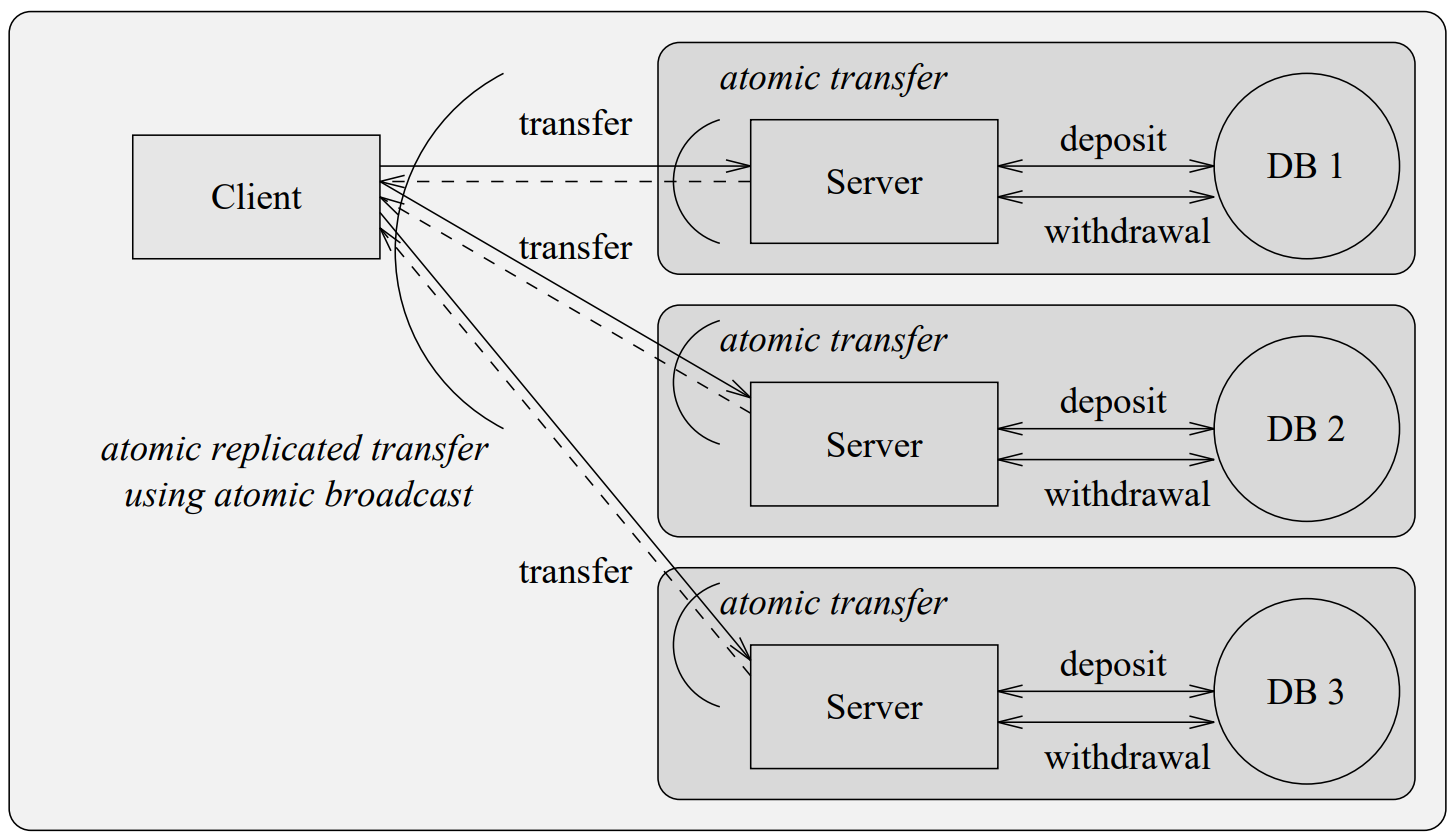}
  \caption{Atomic replicated transfers using atomic broadcasting.}
  \label{atomic}
\end{figure*}

We already started the
implementation of an availability policy using atomic replicated transfers
to provide a sophisticated demonstration of the language
and the power of the toolkit itself.
Fig.~\ref{atomic} illustrates atomic replicated transfers using an atomic
broadcasting feature.

Atomic broadcast can be used within a variety of different applications
where replicated agents that run on different hosts must fulfill a given
task as an atomic action.
That also means that the client must have
a way to distinguish between successful and unsuccessful executions
triggered by broadcasts. We will use version numbers on the raw data
that are manipulated during the remote executions. Using these
version numbers and some atomicity policy, 
the client can decide whether to rollback an execution
on a remote machine or not (see also \cite{Bern87b}). 

An important part of \lokit\ is a constraint-based knowledge broker \cite{pasco/AndreoliBP94}
where agents can proceed even with partial knowledge.
An example application is document merging. Here a merging agent
collects a number of documents that are needed to create a merged
document in some form. However, for timing reasons or other exceptional
behavior, not all the necessary documents are provided in time; the
merging agents must make a decision on the documents ready so far.
Using constraints and making explicit use of partial knowledge, a
new intermediate document state may be reached. Later, when more
documents are available, more knowledge can be incorporated.

Another area of interest concerns security.
Obviously, there are problems arising when every agent in the network
has full access to broadcast information as well as to the activities
of all other agents. This becomes more severe for large hierarchically
structured environments with changing demands for security.  In these
cases the dynamic behavior should be enhanced by appropriate
ask-and-grant mechanisms to restrict the access to agents' private
information.
An example application is remote banking where user and administration people
are distinguished. Both groups of user may have different access rights.
For example, admin people may create/delete accounts whereas normal users
may only access their own accounts to print statement lists, etc.

Furthermore, a debugging facility is needed that shows --
in the best of all worlds -- the dynamic behavior of the concurrent
activities in a graphical way with symbolic notation.
A first step could be to show all agents ready to fire a
broadcast message together with the implied resource allocation 
afterwards.

Finally, \lokit\ will be enhanced to provide linguistic support for asynchronous 
communication such as asynchronous RPCs.



\end{document}